\documentclass[11pt,letterpaper]{article}
\usepackage[top=0.85in,left=0.75in,footskip=0.75in]{geometry}
\usepackage[utf8]{inputenc}
\usepackage{authblk}

\usepackage{graphicx}
\usepackage{amsmath}
\title{Mobility Edge in the Anderson model on partially disordered random regular graphs }
\author[1,*]{O. Valba}
\author[2,3]{A. Gorsky}
\affil[1]{Department of Applied Mathematics, MIEM, National Research University Higher School of Economics, Moscow, Russia}
\affil[2]{Kharkevich Institute for Information Transmission Problems RAS,  Moscow, Russia}
\affil[3]{Moscow Institute for Physics and Technology, Dolgoprudny 141700, Russia}
\affil[*] {ovalba@hse.ru}
\date{}                     
\makeatletter
\renewcommand\@biblabel[1]{#1.}
\makeatother

\begin{document}
\maketitle

\begin{abstract}
In this Letter we study numerically the Anderson model on partially disordered
random regular graphs (RRG) considered as the toy model for a Hilbert space
of interacting disordered  many-body system. The  
protected subsector of  zero-energy states in a many-body system  corresponds 
to clean nodes in RRG ensemble.
Using adjacent gap ratio statistics and IPR we find the sharp mobility edge 
in the spectrum of one-particle Anderson model
above some critical  density of clean nodes. Its position in the spectrum is almost
independent on the disorder strength. The possible application of our result
for the controversial issue of mobility edge in the many-body 
localized (MBL) phase is discussed.
\end{abstract}

\section{Introduction}

Recently new mechanisms of  ergodicity breaking 
in the complicated interacting many-body systems
have been uncovered. The combination of  interaction
and strong enough disorder amounts to the emergent MBL phase 
with full ergodicity breaking \cite{mbl2,mbl1,mbl3,mbl4,mbl5}. 
The Anderson model on RRG serves as the toy 
model for a  identification of MBL phase in the physical space, see
\cite{rev} for the recent review. The many-body localization in the physical space presumably
gets mapped into the one-particle localization in a Hilbert space
\cite{levitov}.
The localization in the Anderson model on the Bethe tree at some critical disorder has 
been established long time ago \cite{thouless} and more 
recently the similar localization on the RRG has been derived  analytically 
\cite{yan} and numerically. This issue has been  analyzed in some details 
in \cite{rrg1,rrg2,rrg3,rrg4,rrg5,rrg6,rrg7,rrg8} and for flat disorder 
$W$ the critical value 
has been established in RRG  for degree $p=3$
with a high accuracy $W_{cr}= 18.17$.

The mechanisms for the destruction of localization are interesting both
from the theoretical and practical viewpoints. In the context of the
destruction of the MBL phase the main mechanism considered involves
a inclusion of  thermal bubbles \cite{dero} which presumably
yield the many-body mobility edge (MBME) 
in interacting many-body systems via the avalanche
mechanism. This issue is 
quite controversial and there are discussions concerning 
the very existence of MBME \cite{mobilityabanin,mob2,mob3,mob4,dero}.
The density of  special states influences
the correlations in  corresponding Hilbert space hopping model \cite{khay1}
and the relation of special states with the MBME was
discussed if the energy \cite{logan} or the density \cite{mobilityabanin}
are considered as the control parameters. It was conjectured that 
the density of thermal bubbles is related to  fractal dimension
of the localized part of the spectrum \cite{tarzia}.

Another quite generic mechanism concerns the effects of the special 
topologically protected degenerated modes. For instance if we assume
that there is no diagonal disorder at all and disorder is present only in the
off-diagonal hopping terms there is the single delocalized mode at 
zero energy which provides the conductivity \cite{dyson,old}. 
Among  the real systems enjoying such mechanism one could mention the
disordered integer QHE in 2d when almost all  modes are localized 
and a conductivity is solely due 
to  zero energy topological modes \cite{2d1,2d2}. Similarly
the topological modes influence the localization in 1d quantum
wires \cite{1d1,1d2}. In both cases the two-coupling renormalization group (RG) flow governs
the evolution of a system from UV to IR. One coupling measures
the strength of disorder while the second one measures the amount of 
topological modes estimated in some manner. 
The RG flow diagram is quite nontrivial and a delocalized 
regime emerges near the boundaries between phases.

Recently the effect of zero modes  has been discussed
for the finite spin systems \cite{zero,zero2,zero3,zero4} in the context of MBL. It was argued that  
several systems  host such exponentially degenerated symmetry protected
modes. In this Letter we shall discuss the combined effect of interaction,
disorder and zero-modes on localization in a real space via the particular toy model in the
Hilbert space.
We shall mimic the presence of the symmetry protected degenerated
zero-modes in a real space by the clean degenerated  nodes
in RRG ensemble. In such toy model 
localization in the interacting many-body system is
described via localization in the  Anderson model on partially disordered  RRG.
The density of the clean zero-energy nodes $\beta$ in a Hilbert space is the
second parameter which quantified the "topological" sector of  Hilbert space. 
It turns out that there is the critical
value of a disorder density $\beta_{cr}$ and
at $\beta < \beta_{cr}$ the one-particle mobility edge 
in the Anderson model on the partially disordered RRG can be identified
unambiguously. It will be confirmed numerically 
by the evaluation of the r-value statistics  and IPR
for the corresponding eigenmodes.

We conjecture that a one-particle mobility edge in the Hilbert space hopping model we found is
related with the MBME in a real space hence 
providing the arguments that  zero-mode degenerated states
play the important role in the MBME formation in disordered system. 
At strong disorder the finite fraction of the protected 
zero modes produces the delocalized states in 
MBL phase in the 
center of  energy band and finally forms  stable 
MBME at large enough density of  protected states.

\section{Model}
We study non-interacting spinless fermions hopping over RRG with connectivity $p=3$ in a potential disorder described by Hamiltonian 
\begin{equation}
H=\sum_{\left<i,j\right>}\left(c_i^+c_j+c_ic_j^+\right)+\sum_{i=1}^{\beta N}\epsilon_ic_i^+c_i,
\label{eq01}
\end{equation}
where the first sum runs over the nearest-neighbor sites of the RRG, the second sum runs over $\beta N $ nodes with potential disorder. The energies $\epsilon_i $ are independent random variables 
sampled from a uniform distribution on $\left[-W/2, W/2\right]$. 
We consider gaps between adjacent levels,
$\delta_i = E_{i+1}-E_i$, where the eigenvalues of a given realization of the Hamiltonian for a given total number of particles, $E_i$, are
listed in ascending order.
The dimensionless quantity we have chosen to
characterize the correlations between adjacent gaps in
the spectrum is the ratio of two consecutive gaps:
\begin{equation}
    r_i = \frac{\min(\delta_i, \delta_{i+1})}{ \max(\delta_i, \delta_{i+1})}.
\label{eq02}
\end{equation}
For uncorrelated Poisson spectrum the probability distribution of this ratio $r$ satisfy by $P (r) = 2/(1 +r)^2$ with the mean
value $\left< r\right>_P = 2 \ln 2-1 \approx 0.386 $. For large Gaussian orthogonal ensemble the mean value $\left< r\right>_{GOE} = \approx 0.53$.

In turn, a direct measure of the (de)localization of the eigenfunctions is obtained by the inverse participation ratio (IPR), $IPR(i) =\sum_n| \psi_n^{(i)}|^4$, where $\psi_n^{(i)}$ is the $i$-th
eigenstate of the matrix and $n$ is the basis state index.
The eigenstate properties are characterized according to
its limiting values as

\begin{equation}
\lim_{N \to \infty} IPR(i)\propto 
 \begin{cases}
   1/N, &\text{delocalization}\\
   const, &\text{localization}
 \end{cases}.
\end{equation}
The IPR of a chaotic eigenstate is $N$-dependent, in contrast to a
localized one.

\section{Numerical results}
We study numerically the dependence of the single-particle localization
on $\beta$ and $W$ using for diagnostics the adjacent gap ratio statistics 
and IPR.

\begin{figure}
\centering
\includegraphics[width=1.0\textwidth]{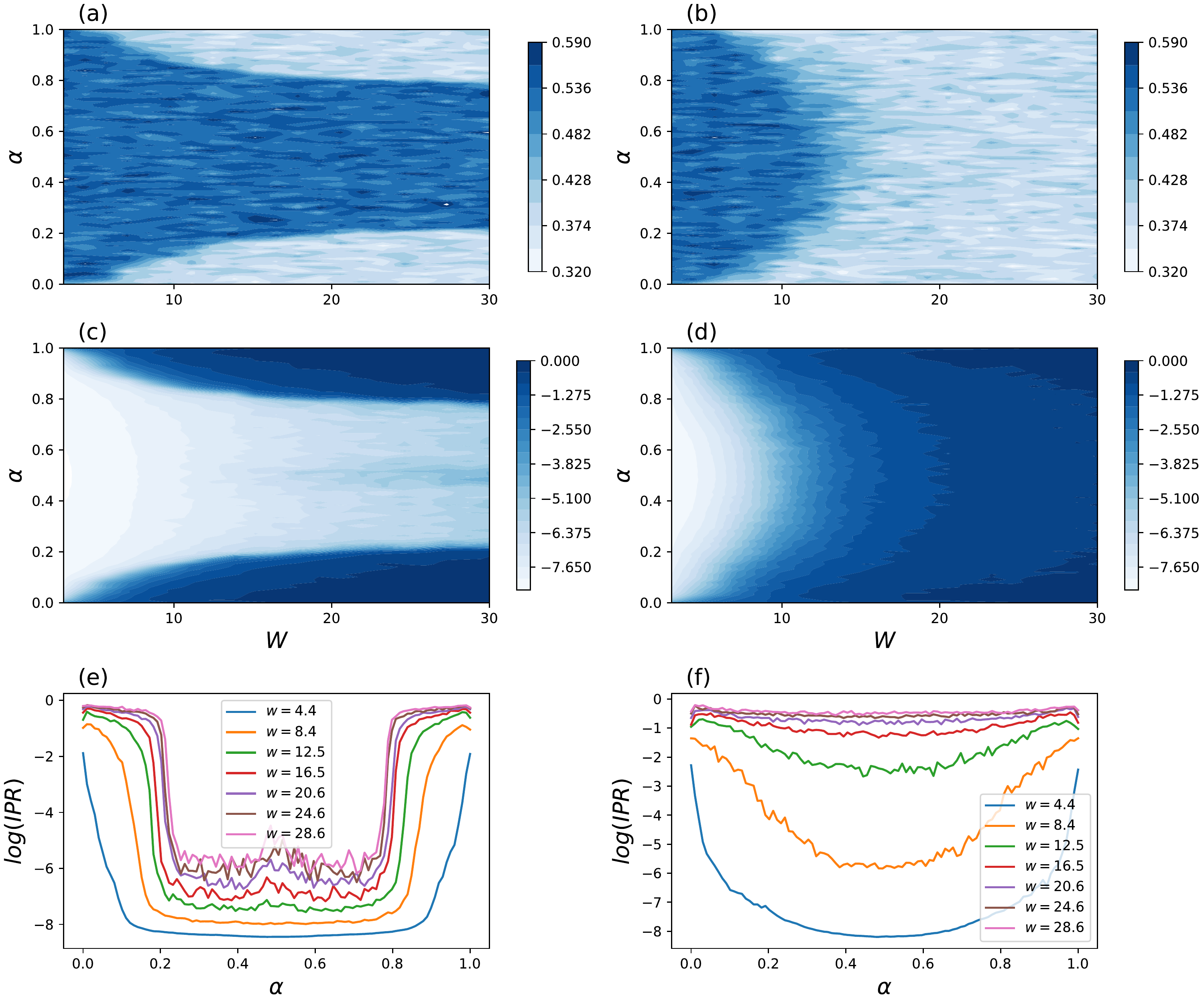}
\caption{The ratio $\left< r\right>$ in dependence  on the disorder value $W$ and the spectrum part $\alpha$ for $\beta=0.5$ (a) and $\beta=1$ (b);the value $\log(IPR)$ in dependence  on the disorder value $W$ and the spectrum part $\alpha$ for $\beta=0.5$ (c) and $\beta=1$ (d); the dependencies of $\log(IPR)$ on the spectrum position $\alpha$ for different values of $W$ and $\beta=0.5$ (e) and $\beta=1$ (f).}
\label{fig01}
\end{figure}

Fig.~\ref{fig01} shows the numerical results for RRG with $N = 16000$ and the degree $p = 3$ and different disorder parameter $\beta$: the left column corresponds to $\beta=0.5$, the right column corresponds to $\beta=1.0$. We analyze the ratio $\left< r\right>$ and IPR for different parts of the spectrum. We divide the sorting spectrum into $k = 100$ equal parts and average the ratio $\left< r\right>$ and IPR over each window. The ordinate $\alpha = i / (N - 1)$ in Fig.~\ref{fig01} corresponds the normalized level position with $i = 0, 1, \dots, N-1,$ the energy level, the ordinate window respectively is $\Delta \alpha = 1 / k$. Fig.~\ref{fig01} (a), (b) demonstrate  the ratio $\left< r\right>$ in dependence on the disorder value $W$ and the spectrum part $\alpha$; (c), (d) presents the results for the values $\log(IPR)$; (e), (f) are respective dependencies of $\log(IPR)$ on the spectrum position $\alpha$ for different values of $W$.

The heat maps Fig.~\ref{fig01} (a), (c) explicitly demonstrate, that there is the mobility edge $\lambda_m$ separating sharply the spectrum into two different regimes for RRG with partial disorder in vertices. For $|\lambda|>\lambda_m $ we observe localisation state with the ratio $\left< r\right>$ close to $\left< r\right>_P$ and independence of IPR on N, while for central spectrum part with $|\lambda|\leq \lambda_m $ the ratio $\left< r\right>$
and IPP indicate on the delocalized state. Note, that the mobility edge $\lambda_m$ weakly depends on the disorder $W$ and is observed even for small $W$. Moreover, we do not observe the phase transition at large $W$ to completely localized phase  which is familiar for completely disordered RRG  (see Fig.~\ref{fig01} (b), (d), (f) with the same plots for $\beta =1.0$).

\begin{figure}
\centering
\includegraphics[width=1.0\textwidth]{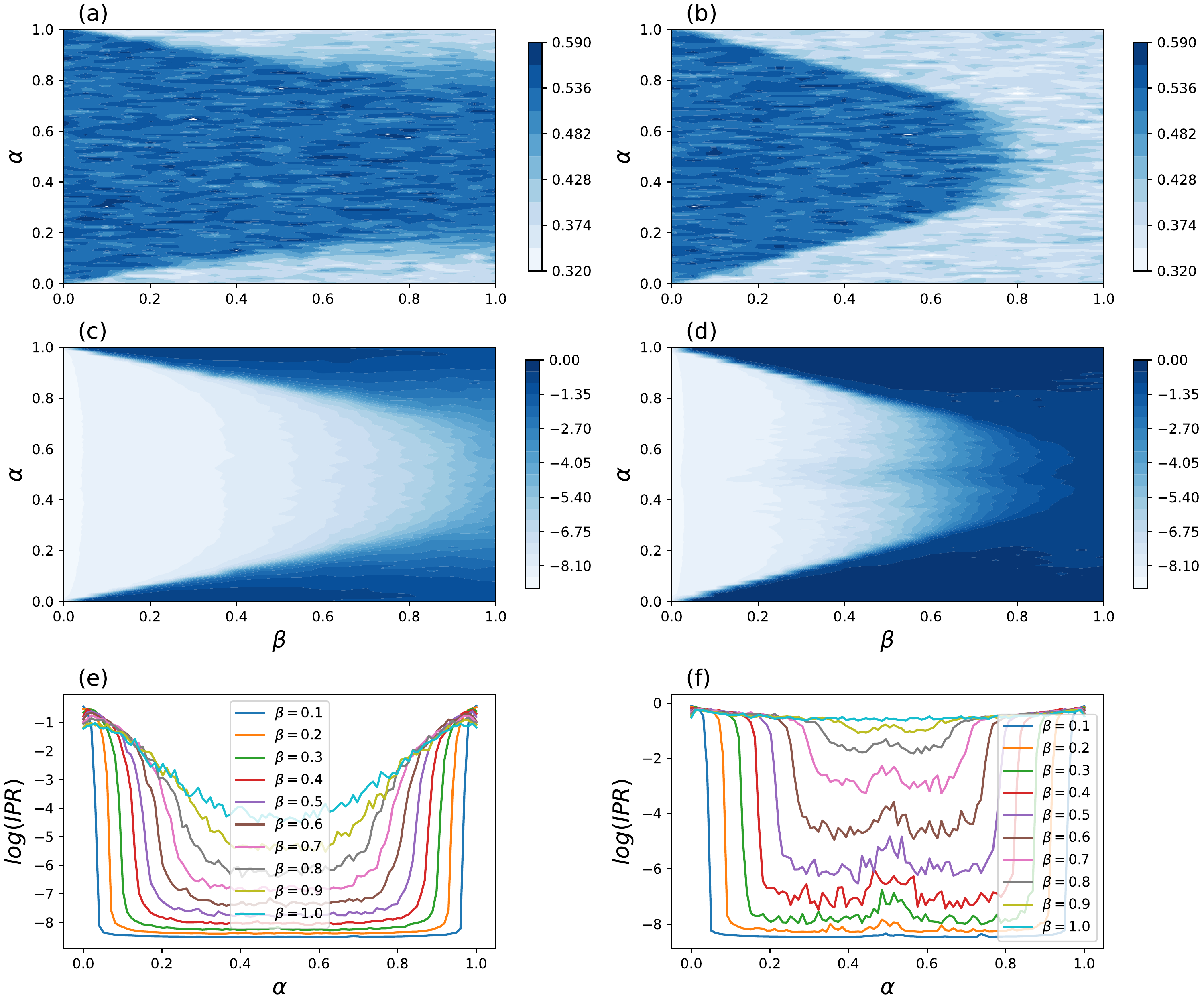}
\caption{The ratio $\left< r\right>$ in dependence  on the disorder parameter $\beta$ and the spectrum part $\alpha$ for $W=10.0$ (a) and $W=25.0$ (b);the value $\log(IPR)$ in dependence  on the disorder parameter $\beta$ and the spectrum part $\alpha$ for $W=10.0$ (c) and $W=25.0$ (d); the dependencies of $\log(IPR)$ on the spectrum position $\alpha$ for different values of $\beta$ and $W=10.0$ (e) and $W=25.0$ (f).}
\label{fig02}
\end{figure}

Fig.~\ref{fig02} demonstrate the numerical results for fixed disorder parameter:the left column corresponds to subcritical disorder parameter $W<W_{cr}$ while the right column corresponds to supercritical value $W>W_{cr}$. In subcritical regime there are the mobility edges for all values $\beta$, while in supercritical regime we observe the existence of the mobility edge just for $\beta<\beta_{cr}$ with $\beta_{cr}=0.8$, while for $\beta>\beta_{cr}$ the dependencies correspond the localization state as for $\beta=1.0$. 

\begin{figure}
\centering
\includegraphics[width=1.0\textwidth]{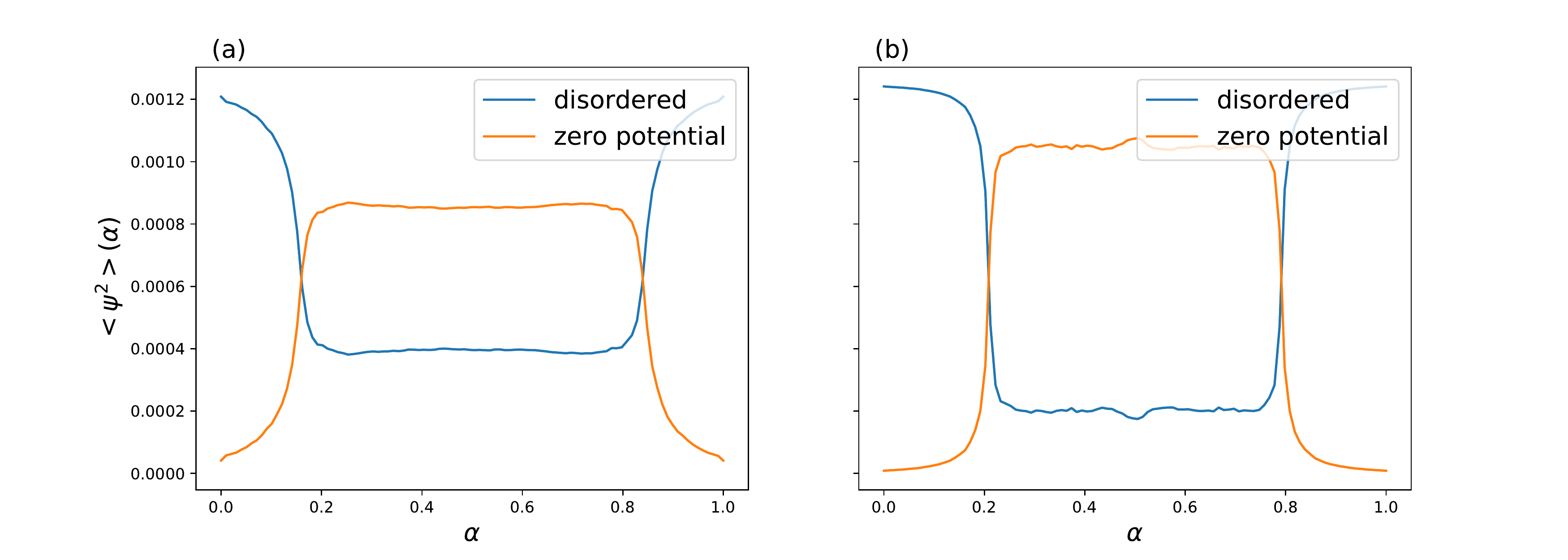}
\caption{The average squared eigenfunction for disordered and zero potential vertices in dependence on  the spectrum part $\alpha$ for $W=10.0$ (a) and $W=25.0$ (b).}
\label{fig03}
\end{figure}

We analyze how eigenfunctions are distributed on network vertices. We calculate squared $i$-th eigenvector component averaging for  the vertices with disorder, i.e, $\left< \psi^2(i)\right > =\frac{1}{\beta N}\sum_{n \in V_d}| \psi_n^{(i)}|^2$ and for the zero potential vertices: $\left< \psi^2(i)\right > =\frac{1}{(1-\beta) N}\sum_{n \in V_0}| \psi_n^{(i)}|^2$,where we denote $V_d$ and $V_0$  the vertices with disorder and with zero potential. Fig.~\ref{fig03} shows the numerical results for the disorder parameter$\beta=0.5$, we plot the value $\left< \psi^2 \right > (\alpha)$, calculated as the average in each of $k$ windows. Let us emphasize once again that the graphs clearly demonstrate the existence of mobility edge, regardless of the disorder level. In addition, we remark that in the localized phase, the eigenfunctions are distributed mainly at the vertices with disorder, while in the delocalized phase, mainly at the clean  vertices.

\section{On the mobility edge in the MBL phase}

In this study we elaborate the combined effect of interaction,disorder and 
the protected states for the many-body system in the toy model of 
partially disordered RRG. Let
us recall the relation between the MBL in the real space and one-particle
localization in the Hilbert space \cite{levitov}. The nodes in the graph
correspond to the states in Fock space in the theory without  interaction. The 
number of nodes equals to the dimension of the Hilbert space and, for
instance, for L spins equals to $2^L$. The matrix elements of the interaction
yield the links between nodes. It was conjectured 
that the one-particle localization in a Hilbert space corresponds to the MBL
phase in the physical space. 

This attractive idea works well qualitatively and provides the explanations
for the several phenomena. For instance, there are strong finite-size effects shifting 
the apparent transition point, exponential growth of the correlation volume when
approaching the transition point, ergodicity of delocalized phase.
On the other hand there are clear-cut arguments which explain that 
the relation between the localization in physical and Hilbert space
can not be exact. For example the tree-like local structure of RRG 
does not fit completely the Hilbert space  of the many-body system.

Having all these reservations in mind one could ask the question if 
the one-particle mobility
edge in the Hilbert space we have found corresponds to a kind of
MBME or another physically
meaningful phenomenon in the real space. 

 To answer this question recall the recent scenario
 concerning the physics in the vicinity of  the mobility edge in MBL.
 It was suggested in \cite{dero} that there are 
 local regions -"thermal bubbles" with  energy density above the
 mobility edge which destroy the localization in a whole
 system via a resonant tunneling. On the other hand 
 it was suggested in \cite{mobilityabanin} that MBME is stable
 with respect to the local regions with high particle
 density which provide the thermal bubbles from \cite{dero}
 and no spreading via the tunneling was observed. The
 expansion of such bubbles into the whole space was
 argued to be more probable origin of delocalization.
 
 Our study suggests that there can be another scenario of 
 MBME formation. Instead of  thermal bubbles the zero-modes
 could do the job and provide the conductivity in the mid
 of the spectrum. We do not expect that  zero-modes are  
 completely protected under the disorder and interaction.
 More relevant scenario involves the formation of  non-vanishing
 density of the quasi-zero modes in the mid of the 
 spectrum somewhat similar to the situation in
 QCD although one has to have in mind a lot of 
 reservations concerning this analogy. Indeed we have found
 numerically that the states in the delocalized mid part
 of the spectrum involve not only the clean modes but also
 the disorder dependent admixture of the dirty modes. Therefore
 they propagate mainly along the clean nodes with some 
 disorder dependent contribution of regions with dirty states.
 
 In QCD the localization properties of the Dirac
 operator eigenfunctions in 4d Euclidean space-time are considered.
 In the confinement phase the physical many-body spectrum of QCD  involves the 
 composite mesons and baryons only. It is well-known that the density of quasi-zero modes $\rho(0)$ of the
 Dirac operator yields the 
 breakdown of the chiral symmetry in the confinement phase via formation of the
 chiral condensate according to the Casher-Banks relation
 $<\bar{\Psi}\Psi>= V^{-1}\pi \rho(0)$,  where V is the 4-volume.
 The Dirac operator enjoys the discrete symmetry  $[H_{D}^{4d},\gamma_5]=0$
 hence its spectrum is symmetric under $E\rightarrow -E$.
 All eigenmodes of the Dirac operator are delocalized in confined
 phase \cite{lattice} which usually is explained via the 
 instanton liquid model for QCD ground state. The delocalization
 is attributed to the overlap of the topological fermionic zero modes 
 localized on the
 instantons and antiinstantons. Moreover the lattice 
 studies indicate that a delocalization occurs 
 not in full 4D Euclidean space but only along some 
 lower dimensional sub-manifolds
 somewhat similar to the fracton picture.
 
 When we increase the temperature the deconfinement phase transition
 occurs at some critical $T_c$. At $T>T_c$ the spectrum of QCD 
 involves the interacting quarks and gluons. Remarkably exactly
 at $T=T_c$ the mobility edge in the Dirac operator spectrum gets emerged \cite{lattice}.
 It can be related to the formation of black hole horizon in the holographic
 approach \cite{gl}. In the deconfined phase two essential phenomena
 happen. First, the disordered condensate of the Polyakov lines gets
 formed providing the strong enough disorder. Secondly the density
 of the topological defects - instantons gets decreased due to
 the asymptotic freedom since 
 it  is proportional to $\exp(-\frac{1}{g_{YM}^2(T)})$. Hence indeed
 like in our model we have three characteristics which influence
 a localization: the strength of interaction, the density of 
 topological defects and the strength of disorder. Like in our 
 model when the topological defects in the QCD confinement phase dominate we have complete
 delocalization in  one-particle spectrum while if 
 the disorder becomes strong enough 
 the mobility edge gets emerged in the deconfined phase .

\section{Discussion}

In this Letter we consider partially disordered RRG 
as the toy model of a Hilbert space for some
interacting disordered many-body system with the 
topologically protected subsector.
The nodes of RRG free from disorder correspond to 
zero-energy  protected states in
many-body system. To some extend our model probes
the effects of disorder on  zero-mode states.

It is found that at some density 
of  clean nodes in partially disordered RRG the sharp mobility
edge emerges in the spectrum of Anderson model and exists up 
to arbitrarily large diagonal
flat disorder $W$. We have studied the distribution 
of the eigenfunctions in RRG and have found that  
localized states are distributed almost solely within the dirty
nodes while the delocalized part of the spectrum 
mainly involves the clean nodes with small disorder
dependent contribution of the dirty nodes.

The model certainly oversimplifies the
issue  nevertheless it can be considered as the
indication that a one-particle mobility edge 
in the hopping model in the Hilbert space
and a mobility edge in the MBL phase could be related.
Indeed if the mechanism behind the  mobility edge in MBL 
involves density of highly degenerate zero-modes in the physical space 
then the density of clean nodes in the Hilbert space
is its relevant counterpart. It would be important to get
the analytic description of the observed phenomena using
approach developed in \cite{yan,analytic3}.

The mobility edge emerges in RRG ensemble also at another
occasion when a chemical potential for the 3-cycles
is added. At some critical value of the chemical
potential the RRG with node degree $p$ gets fragmented into the $\frac{N}{p}$
clusters which are almost complete graphs  \cite{defragmentation}.
The corresponding spectral density develops the second
non-perturbative band filled by eigenvalue instantons
corresponding to clusters.
The emerging mobility edge separates the continuum filled by
the delocalized modes and the second non-perturbative band with  
localized modes obeying Poisson statistics \cite{localization}.
The formation of clusters is suppressed by factors $e^{-N}$
however they do not "evaporate" completely at large $N$
yielding the star-like hubs \cite{hubs}. The similar 
fragmentation of the RRG takes place also if the chemical potential
for 4-cycles is added \cite{trug,gv2}. In this case the emerging
bipartiteness of the fragments is new peculiar feature of the model \cite{gv2}.

It is natural to compare  this
pattern with other mechanisms of Hilbert space
fragmentation which influence the localization,
in particular quantum many-body scars (QMBS), see \cite{bernevig2,scarreview,papic}.
The scars are related to the symmetries of the many-body Hamiltonian
 and are protected algebraically when the interaction
 does not ruin the symmetry completely \cite{mori,abaninscars,klebanov,scar1,scar2}. 
Note that  QMBS are not 
degenerated enough to fill the finite part of the Hilbert space.
It is assumed that  scars
are quantum counterpart of the peculiar unstable semi-classical orbits
in  classically chaotic systems. Somewhat similarly the
k-cycles in the Hilbert space which induce fragmentation of RRG
presumably correspond to the
k- resonanses which also are quantum counterpart of
the peculiar semi-classical orbits \cite{basko}. It would be interesting
to discuss their possible relations.
It would be also
interesting to combine the effects of a diagonal disorder
and chemical potentials for k-cycles in RRG representation
of the Hilbert space together.
It could be the toy model for a interplay of  scars with
disorder discussed in \cite{vavilov}.

We are grateful to A. Kamenev and I. Khaymovich for the useful discussions.
The work  was supported by the Russian Science Foundation grant 21-11-00215.


\end{document}